\documentclass[aps, prl, twocolumn,groupedaddress]{revtex4}
\usepackage{graphicx}
\usepackage{amsmath}
\usepackage{amssymb}

\newcommand{\bb}{\begin{equation}}
\newcommand{\ee}{\end{equation}}
\newcommand{\ba}{\begin{eqnarray*}}
\newcommand{\ea}{\end{eqnarray*}}
\newcommand{\rhor}{\rho({\bf r})}
\newcommand{\dd}{{\rm d}}
\newcommand{\rr}{{\mathbf r}}
\newcommand{\dr}{{\rm d}{\bf r}}

\usepackage{graphicx}
\usepackage{dcolumn}
\usepackage{bm}
\usepackage{longtable}

\begin{document}



\title{Does adsorption in a single nanogroove exhibit hysteresis?}

\author{Alexandr \surname{Malijevsk\'y}} 

\affiliation{{Department of Physical Chemistry, Institute of
Chemical Technology, Prague, 166 28 Prague 6, Czech Republic }\\
E. H{\'a}la Laboratory of Thermodynamics, Institute of Chemical Process Fundamentals, Academy of Sciences, 16502 Prague 6, Czech Republic}

\begin{abstract}
A simple fluid, in a microscopic capillary capped at one end, is studied by means of fundamental measure density functional.
The model represents a single, infinitely long nanogroove with long-range wall-fluid attractive (dispersion) forces. It
is shown that the presence or absence of hysteresis in adsorption isotherms is determined by wetting properties of the wall as follows: Above wetting temperature, $T_w$, appropriate to a single wall of the groove, the adsorption is a continuous process
corresponding to a rise of a meniscus from the capped to the open end of the groove. For a sufficiently deep capillary the meniscus rise is shown to be a steep,
yet continuous process taking place near the capillary condensation of a corresponding slit. However, for temperatures lower than $T_w$ the condensation exhibits
a first-order transition accompanied by hysteresis of the adsorption isotherm. Finally, it is shown that hysteresis may occur even for $T>T_w$ as a
consequence of prewetting on the side and bottom walls of the groove.
\end{abstract}

\pacs{68.08.Bc, 05.70.Np, 05.70.Fh}
\keywords{Capillary condensation, Density functional theory, Fundamental measure theory, Lennard-Jones, Wetting, Adsorption.}

\maketitle

The interfacial properties of fluids in confining geometries play a key role in several branches of physics and chemistry, and are vital for numerous engineering
applications, see, e.g., Ref \cite{gelb1}.
Two of the most fundamental interfacial phenomena involve a development of a liquid phase on a single wall and between two parallel \emph{unbounded} walls at  a
pressure below the phase coexistence. The first, known as complete wetting, is characterized by a growth of an adsorbed liquid film
$\ell\sim\Delta\mu^{-\beta_s^{co}}$. Here $\ell$ is the film thickness and $\Delta\mu$ is a deviation of the chemical potential from its saturation value, $\mu_{\rm
sat}(T)$, above a wetting temperature $T_w$ (which corresponds to a zero contact angle). The exponent $\beta_s^{co}$ is non-universal and depends on the character of molecular interaction; in particular,
$\beta_s^{co}=0$ for short-ranged forces, while $\beta_s^{co}=1/3$ when dispersion forces are involved \cite{dietrich}. The second phenomenon, capillary condensation, differs from the
previous one in two respects. First, it is a first-order transition, and, second, it reflects the finite size shift of the (3D) bulk phase boundary rather than surface phenomena at the wall.  For a non-retarded van der Waals fluid-wall interaction and in the limit of a large distance $L$ between the walls,
the modified Kelvin equation \cite{derj, evmar} predicts the undersaturation at which the capillary condensation occurs at $T>T_w$:
 \bb
 \Delta\mu(L)=\mu_{\rm sat}-\mu_{\rm cc}(L)\approx \frac{2\gamma}{(\rho_l-\rho_g)(L-3\ell)}\,.\label{kelvin}
 \ee
Here $\gamma$ is the surface tension of a free liquid-gas interface,  $\rho_l$ and $\rho_g$ are densities of the coexisting bulk phases, and $\ell$ is an
adsorbed-film thickness. It should be noted that the details of the intermolecular forces in (\ref{kelvin}) are reflected only in the Derjaguin's correction
(the factor of three multiplying the film thickness).

While adsorption phenomena in systems that can be described by a one-dimensional density distribution are rather well understood, interfacial phenomena on patterned surfaces, which are characterised by a variation in the density distribution at least in 2D, have recently become the focus of considerable interest, see, e.g., Ref.
\cite{bonn}. The presence of a complex distribution of possible substrate topographies introduces a challenging task of relating geometrical and well-defined substrate models to the thermodynamics of fluid adsorption. Models suited to this task include linear wedges, cones, grooves, and pitted surfaces to name a few of the
most popular \cite{nature, fil, fil2, cov, tas, marconi, darbellay, evans_cc}.
The interest in this research has been largely motivated by new advances in nanofabrication techniques \cite{mistura} which made possible a direct comparison
between theoretical predictions and experiment \cite{hofmann}. Apart from the theoretical interest, the adsorption in nanopatterned surfaces has attracted considerable attention due to its application in micro- and nanofluidics \cite{microfluidics}.

The purpose of the paper is to give a microscopic description of the adsorption of a simple fluid
in a single, infinitely long groove of finite depth. In particular,
we wish to know whether the adsorption isotherms in this model exhibit hysteresis or not. The main conclusion based on earlier studies \cite{darbellay, evans_cc,
gelb, marconi, roth} is that in the case of complete wetting the hysteresis vanishes due to the presence of the bottom wall. Here we present a detailed microscopic study of this issue and discuss the effect of the wetting properties on the character of the groove-adsorption.



Consider a semi-infinite solid slab with a uniform one-body density $\rho_w$ spanning a domain $\mathbb{S}=\mathbb{R}\otimes\mathbb{R}\otimes(-\infty,L_z),\;L_z>0$, in 3D Cartesians. Imagine
that an infinitely long groove of width $L_x$ and depth $L_z$, occupying a subspace $\mathbb{G}=(0,L_x)\otimes(-\infty,\infty)\otimes(0,L_z)$, is sculpted into the slab. We assume that the groove is subject
to a potential:
 \bb
 V(x,z)=\left\{\begin{array}{ll} 0\,,&x<\sigma_w\;{\rm or}\;x>L_x-\sigma_w\;{\rm or}\;z<\sigma_w\,;\\
\tilde{V}(x,z)\,,& {\rm elsewhere\,,}\end{array}\right.
 \ee
 where $\tilde{V}(x,z)=\rho_w\int_\mathbb{W}\phi_w(|\rr-\rr'|)\dr'$, $\mathbb{W}=\mathbb{S}\setminus\mathbb{G}$, and
 \bb
 \phi_w(r)=-4\varepsilon_w\left(\frac{\sigma_w}{r}\right)^{6}\,.
 \ee
Thus, we assume that the fluid particles interact with the substrate atoms via long-ranged (dispersion) forces characterized by the parameters $\varepsilon_w$
and $\sigma_w$.

After integration, $\tilde{V}(x,z)$ can be expressed as follows:
 \bb
  \tilde{V}(x,z)=V_{1}(z)+V_{2}(x,z)+V_{2}(L_x-x,z)\,,\label{V}
\ee
 with
$
 V_{1}(z)=\frac{2\alpha_w}{z^3}
$
 and
$
 V_{2}(x,z)=\alpha_w\left[\psi(x,z)+\psi(x,L_z-z)\right]\,,
$
 where we have defined
$
\alpha_w=-\frac{1}{3}\pi\varepsilon_w\rho_w\sigma_w^6
$
and
$$
\psi(x,z)=\frac{2z^4+x^2z^2+2x^4}{2x^3z^3\sqrt{x^2+z^2}}-\frac{1}{z^3}\,.
$$
The fluid-fluid interaction is given by
 \bb
 \phi(r)=\left\{\begin{array}{ll} \infty,& r<\sigma\,;\\
-4\varepsilon\left(\frac{\sigma}{r}\right)^{6},&\sigma<r<r_c\,;\\
0,& r>r_c\,.\end{array}\right. \label{phi}
 \ee
In the following, parameters $\sigma$ and $\varepsilon$ will be used as length and energy units, respectively, and the cutoff is set to $r_c=2.5\,\sigma$. The wall parameters are fixed to $\sigma_w=\sigma$ and
$\varepsilon_w=1.2\,\varepsilon$. The wetting temperature for such a model is $T_w=0.83\,T_c$, where $k_BT_c/\varepsilon=1.41$ is the critical temperature.

Within a classical density functional theory \cite{evans_dft}, the equilibrium density profile is found by minimizing the grand potential functional
 \bb
 \Omega[\rho]={\cal F}[\rho]+\int\dd\rr\rhor[V(\rr)-\mu]\,,\label{om}
 \ee
where $\mu$ is the chemical potential. ${\cal F}[\rho]$ is the intrinsic free energy functional of the one-body density, $\rhor$,
whose excess (over ideal gas) part is treated as a perturbation about a hard-sphere reference fluid:
  \bb
  {\cal F}_{\rm ex}[\rho]={\cal F}_{\rm hs}[\rho]+\frac{1}{2}\int\dd\rr\rhor\int\dd\rr\rho(\rr')\phi(|\rr-\rr'|)\,. \label{f}
  \ee
The hard sphere part of the excess free energy functional is approximated by means of the Rosenfeld fundamental measure theory (FMT) \cite{ros}
 \bb
 {\cal F}_{\rm hs}[\rho]=\frac{1}{\beta}\int\dd\rr\Phi(\{n_\alpha\})\,,
 \ee
where $\Phi$ is a function of the weighted densities $\{n_\alpha(\rr)\}$, and $\beta=1/k_BT$.
%
Numerical calculations were carried out on a 2D grid with mesh $0.05\,\sigma$.

One should note that the functional (\ref{f}) with ${\cal F}_{\rm hs}[\rho]$ obtained from the FMT is known to properly account for the short-ranged correlations
that play a significant role in confined systems. In our nanogroove model one expects strong packing effects at the vicinity of the walls (as in the case of
common slit models) but additional and particularly strong inhomogeneities are also expected at the edges of the groove. Finally, our DFT is thermodynamically
consistent and yields the correct divergence of wetting film thickness for systems that exhibit dispersion forces. All these properties are of key importance for
a reliable description of the phase behaviour of the model.

\begin{figure}[htp]
\includegraphics[width=0.35\textwidth]{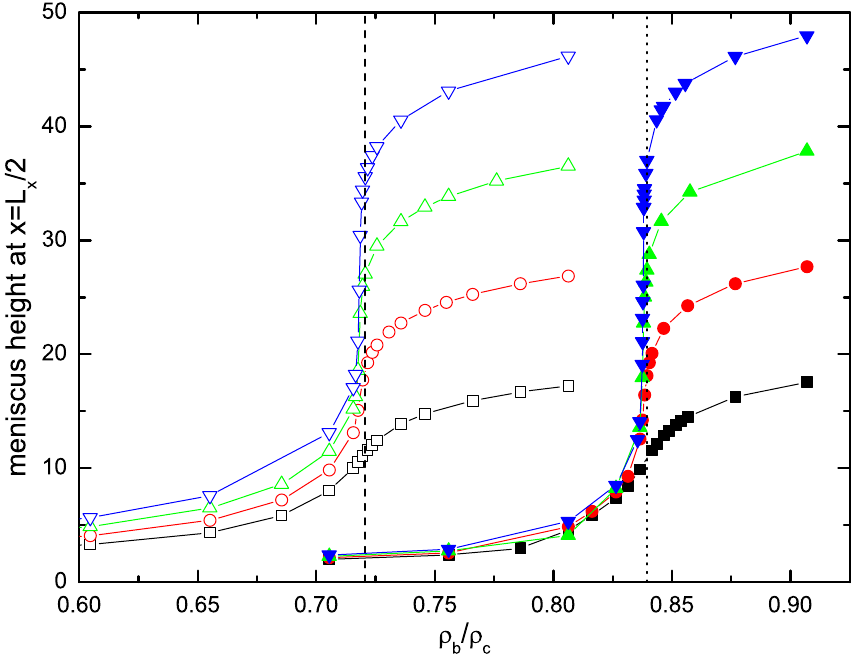}
\caption{Meniscus rise in a groove of width $L_x=7\,\sigma$ (empty symbols) and $L_x=12\,\sigma$ (full symbols) as a function of undersaturation ($\rho_b$ is the bulk density, $\rho_c$ is the critical density). The depths of the grooves are from the bottom $L_z/\sigma=20, 30, 40$, and $50$. The curves are the guides to the
eye. The vertical lines denote densities corresponding to capillary condensation in an infinite slit of a width $L_x=7\,\sigma$ (dashed) and $L_x=12\,\sigma$ (dotted). For $T=0.92\,T_c$.}\label{fig1}
\end{figure}

In the following, we summarize the most-important features of adsorption in the model groove that can be inferred from our DFT calculations. We start by
considering temperature $T=0.92\,T_c$, which is above the wetting temperature, $T_w$, of a corresponding planar wall. In such a case, a meniscus separating a
gas-like and a liquid-like phase is formed near the bottom end, and continuously rises upon increasing the chemical potential towards the two-phase coexistence region.
Position of the meniscus can be defined as $\ell(x)\equiv\int\dd z(\rho(x,z)-\rho_b)/\Delta\rho$, where $\Delta\rho=\rho_l-\rho_g$ is the difference between liquid
and vapour densities at the bulk coexistence. In Fig. 1 the meniscus rise is shown for grooves of $L_x=12\,\sigma$ and $L_x=7\,\sigma$ and several depths. As is
apparent, the groove filling is a continuous process, which should be contrasted to capillary condensation in infinite slits. Furthermore, as $L_z$ is increased,
the meniscus rise becomes steeper and steeper upon approaching the capillary condensation indicated by the vertical lines. Such results should be compared with a
schematic plot (Fig. 1) in reference \cite{evans_cc}.

\begin{figure}[htp]
\includegraphics[width=0.35\textwidth]{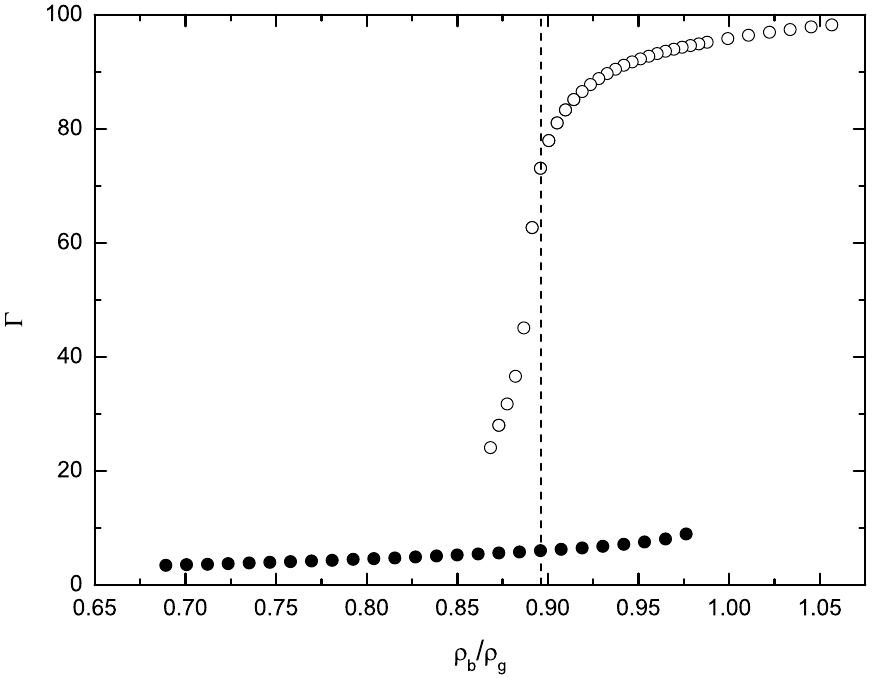}
\caption{Adsorption isotherm at temperature $T=0.81\,T_c$ in a groove of width $L_x=12\,\sigma$ and depth $L_z=20\,\sigma$;
$\rho_b$ is the bulk density, $\rho_g$ is the density of a saturated gas. Full symbols denote the adsorption path, empty symbols denote the desorption path. The vertical dashed
line denotes a density where two distinct states coexist.} \label{fig2}
\end{figure}

Next consider $T=0.81\,T_c$, which is slightly below $T_w$. Now, the scenario of the fluid adsorption dramatically changes. As displayed in Fig. 2, the system exhibits a hysteresis in adsorption per unit length
$\Gamma=\int_\mathbb{G}\rho(x,z)\,\dd x\, \dd y$, indicating a first order transition resembling capillary condensation in infinite slits. In order to understand the different behaviour of the adsorption in the
two cases, let us discuss the mechanism of the groove filling for $T>T_w$ and $T<T_w$.

\begin{figure}[htp]
\includegraphics[width=0.2\textwidth]{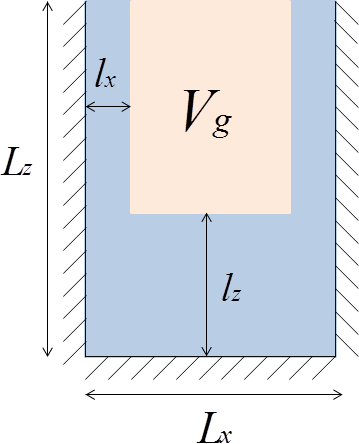}
\caption{Schematic illustration of the slab model.}\label{fig3}
\end{figure}

For $T>T_w$, the groove filling
can be described with the help of a simple slab model, as schematized in Fig. 3. Substituting a sharp-kink approximation for the density profile in (\ref{om}),
the excess (over the groove completely filled with a liquid) grand potential:
 \begin{eqnarray}
  &&\frac{\Omega^{\rm ex}(\ell_x,
\ell_z)}{L}=(p_l^+-p)(L_x-2\ell_x)(L_z-\ell_z)\\
&&+\gamma[2(L_z-\ell_z)+(L_x-2\ell_x)]+\frac{\rho_g-\rho_l}{L}\int_{V_g}V(\rr)\dr\,,\nonumber \label{om_ska}
 \end{eqnarray}
where $p_l^+$ is the bulk pressure of the metastable liquid and $p$ is that of the gas reservoir.
When minimized with respect to $\ell_z$ (minimization wrt to $\ell_x$ reproduces Eq.~(\ref{kelvin})), one obtains in the limit of
large $L_z$:
\bb
\Delta p+\frac{A}{\ell^4}+{\cal O}(1/\ell_z^6)=0\,,\label{delta_p}
\ee
where $\Delta p=p-p_l^+-\frac{2\gamma}{L_x-3\ell_x}$ and $A=\frac{9}{8}\frac{\alpha_w(\rho_l-\rho_g)L_x(L_x-2\ell_x)}{L_x-3\ell_x}$.
Comparing (\ref{delta_p}) with (\ref{kelvin}) and assuming that $\Delta\mu(L_x)$ is small, the meniscus rise satisfies (cf. Ref. \cite{evans_cc})
 \bb
 \ell_z\sim(\mu_{\rm cc}(L_x)-\mu)^{-1/4}\,.\label{fil}
 \ee
The same conclusion was obtained in Ref. \cite{evans_cc}. However, in Ref. \cite{evans_cc} the authors ignored the influence of $\ell_x$, while the connection with Eq. (\ref{kelvin}) has been made more explicit here.
Thus, for $T>T_w$ one observes a continuous filling of a groove, similar to a complete wetting on a planar wall. However, the differences with the latter are:
i) the deep groove becomes filled in the limit $\mu\rightarrow\mu(L_x)$ rather than for $\mu\rightarrow\mu_{\rm sat}$; and
ii) the critical exponent of the groove filling is $1/4$, i.e., the process is somewhat slower than the complete wetting.

In contrast, for $T<T_w$ only a microscopically thin film is formed on a free wall.
Therefore, the system behaves in a qualitatively same fashion as in an open slit and undergoes a first-order transition accompanied by pronounced hysteresis.
We emphasize that Eq.~{\ref{om_ska}) cannot be applied anymore in this regime, since inclusion of higher order terms due to repulsive interactions would be needed. These would give a global minimum of $\Omega^{\rm ex}$ at a finite distance of $\ell_z$.

Finally, let us consider temperature $T=0.91\,T_c$. Such a temperature is again above $T_w$ but this time (slightly) lower than $T_{cs}$, the prewetting critical temperature corresponding to a single wall. It is well known
\cite{prew} that the prewetting transition between a thin and thick layers on a single wall may also occur in slits, but the phenomenon is preceded by capillary condensation unless the separation between the walls is large. It is
interesting to examine how prewetting is reflected in our model groove in the absence of capillary condensation (above $T_w$). In Fig.~4 an adsorption isotherm for a model $L_x=L_z=50\,\sigma$ is depicted. We observe that
the otherwise-continuous increase in adsorption, due to meniscus rise, exhibits two small van der Walls loops, both slightly below the bulk density appropriate to prewetting. The first of the two loops corresponds to a discontinuous
jump in the meniscus height, reflecting prewetting on the bottom wall, cf. Fig. 5 (top). Such a transition is shifted to a higher undersaturation compared to prewetting on a single wall, $\rho_b^{\rm pw}/\rho_{\rm sat}$, due to
the influence of the side walls that decreases the potential inside the groove. The second loop corresponds to prewetting on the side walls, cf. Fig. 5 (bottom). Again, this transition is shifted below $\rho_b^{\rm
pw}/\rho_{\rm sat}$ but the difference is smaller, since the effect of the bottom wall on the vertical liquid films is relatively weak (for large $z$ the potential of the bottom wall decays as $z^{-3}$). However, the scenario
may be different and the sequence of the two transitions be reversed if the side walls are stronger adsorbents than the bottom wall.

\begin{figure}[htp]
\includegraphics[width=0.33\textwidth]{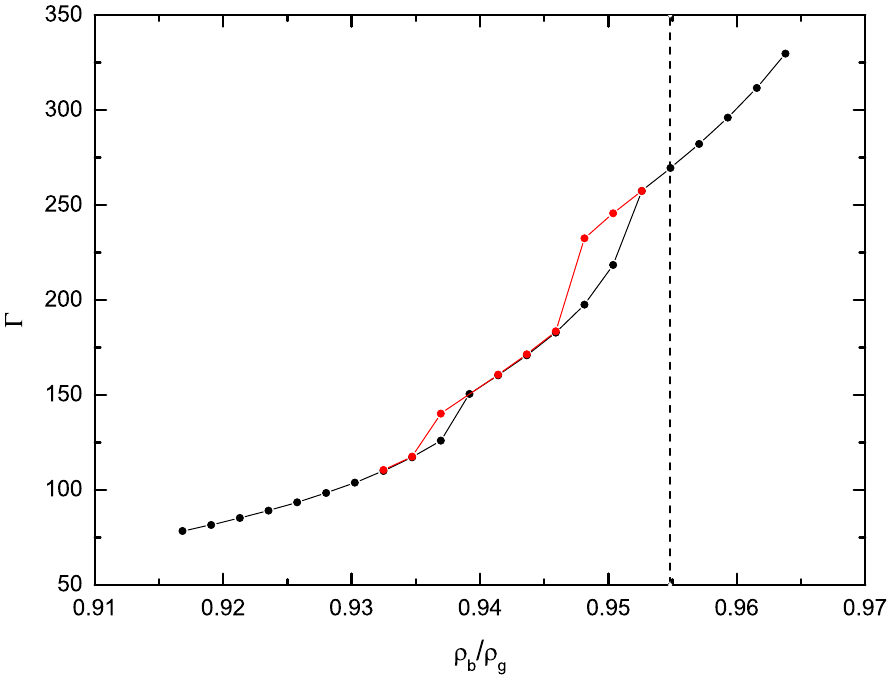}
\caption{Adsorption isotherm at temperature $T=0.91\,T_c$ in a groove of width $L_x=50\sigma$ and depth $L_z=50\,\sigma$. Black symbols denote the adsorption path, red symbols denote the desorption path. The lines are the
guides to the eye. The vertical dashed line denotes a density corresponding to the prewetting transition in an open slit of width $L=50\sigma$.} \label{fig2}
\end{figure}

\begin{figure}[htpb]
\includegraphics[width=0.22\textwidth]{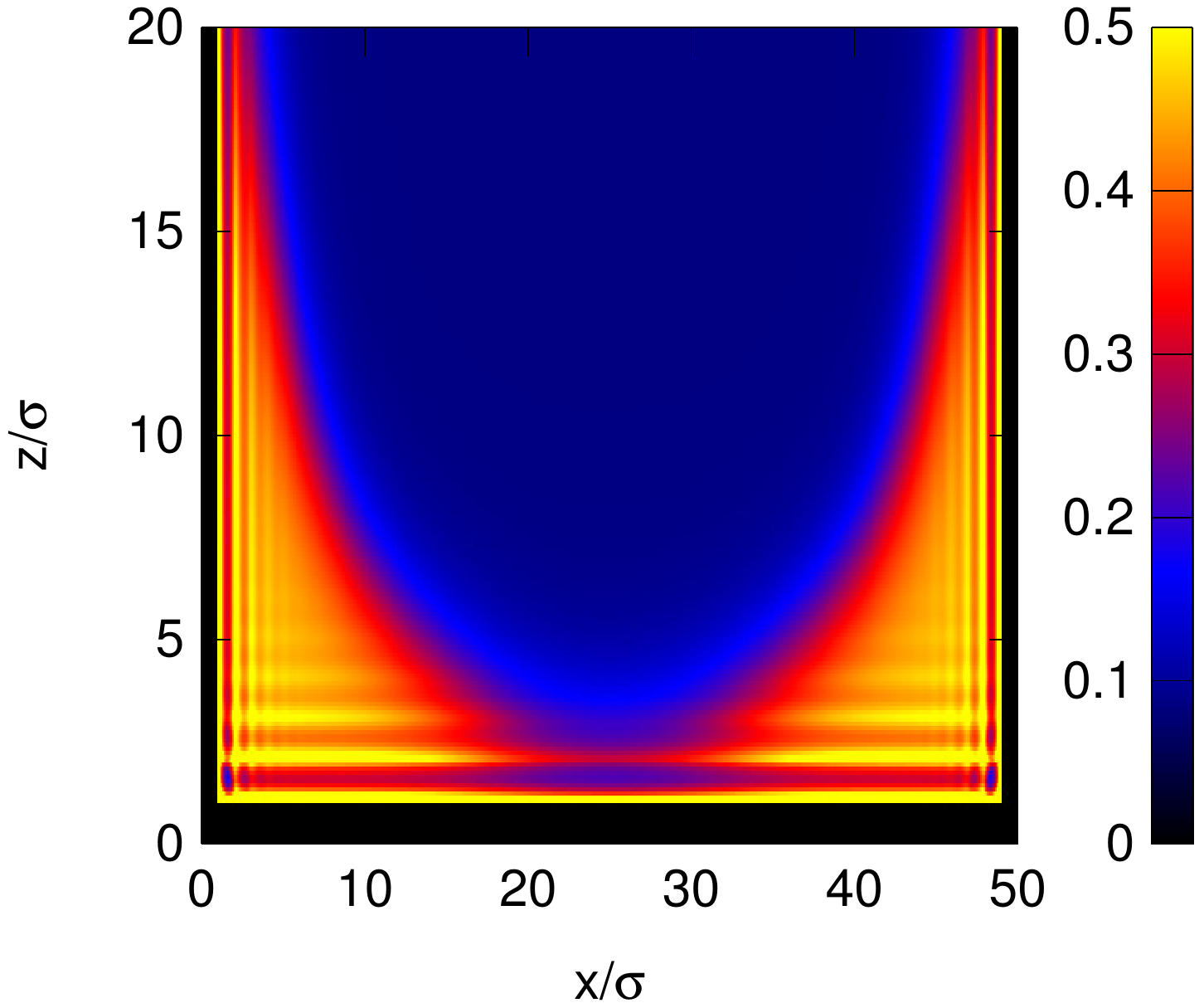} \includegraphics[width=0.22\textwidth]{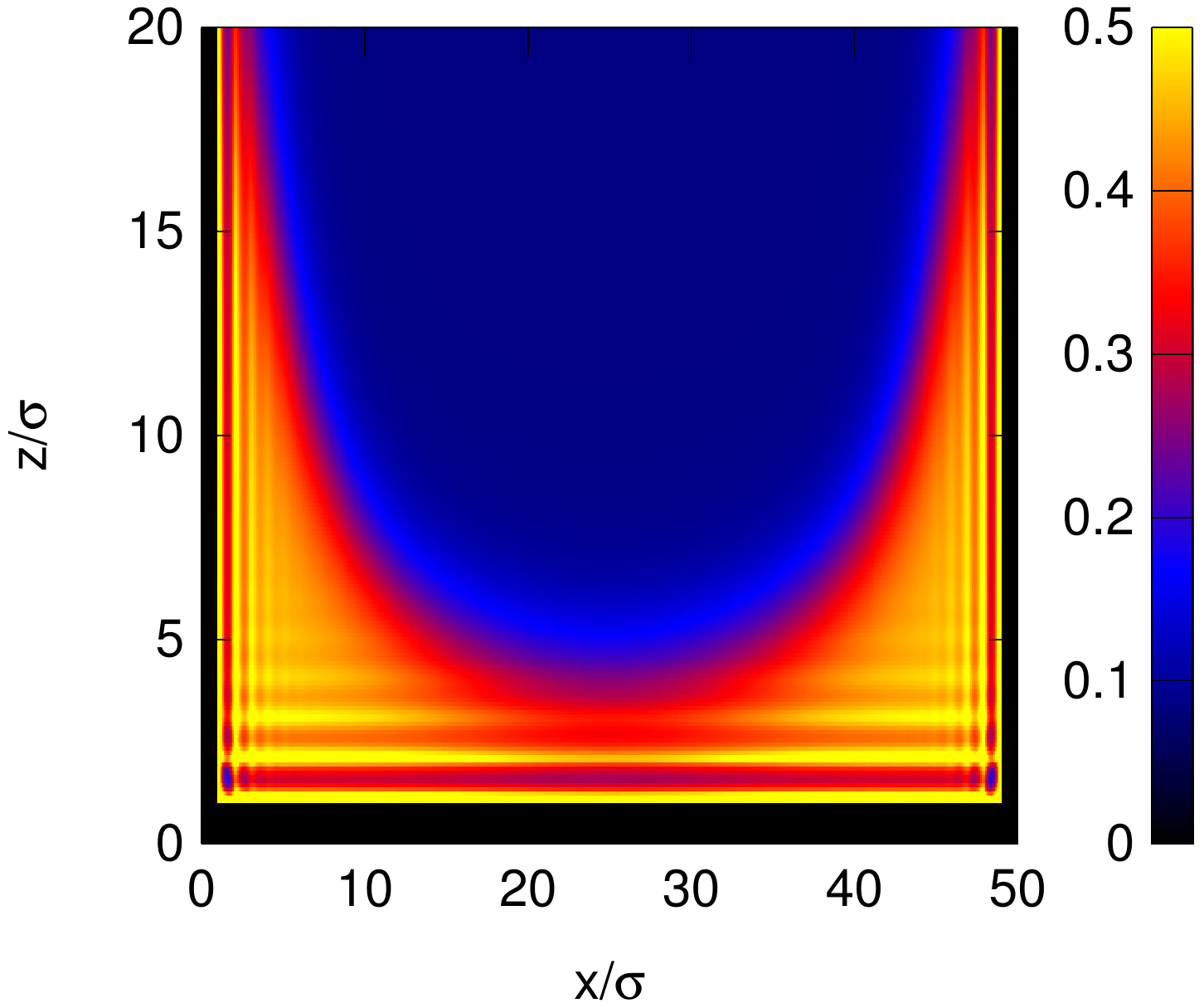}
\includegraphics[width=0.22\textwidth]{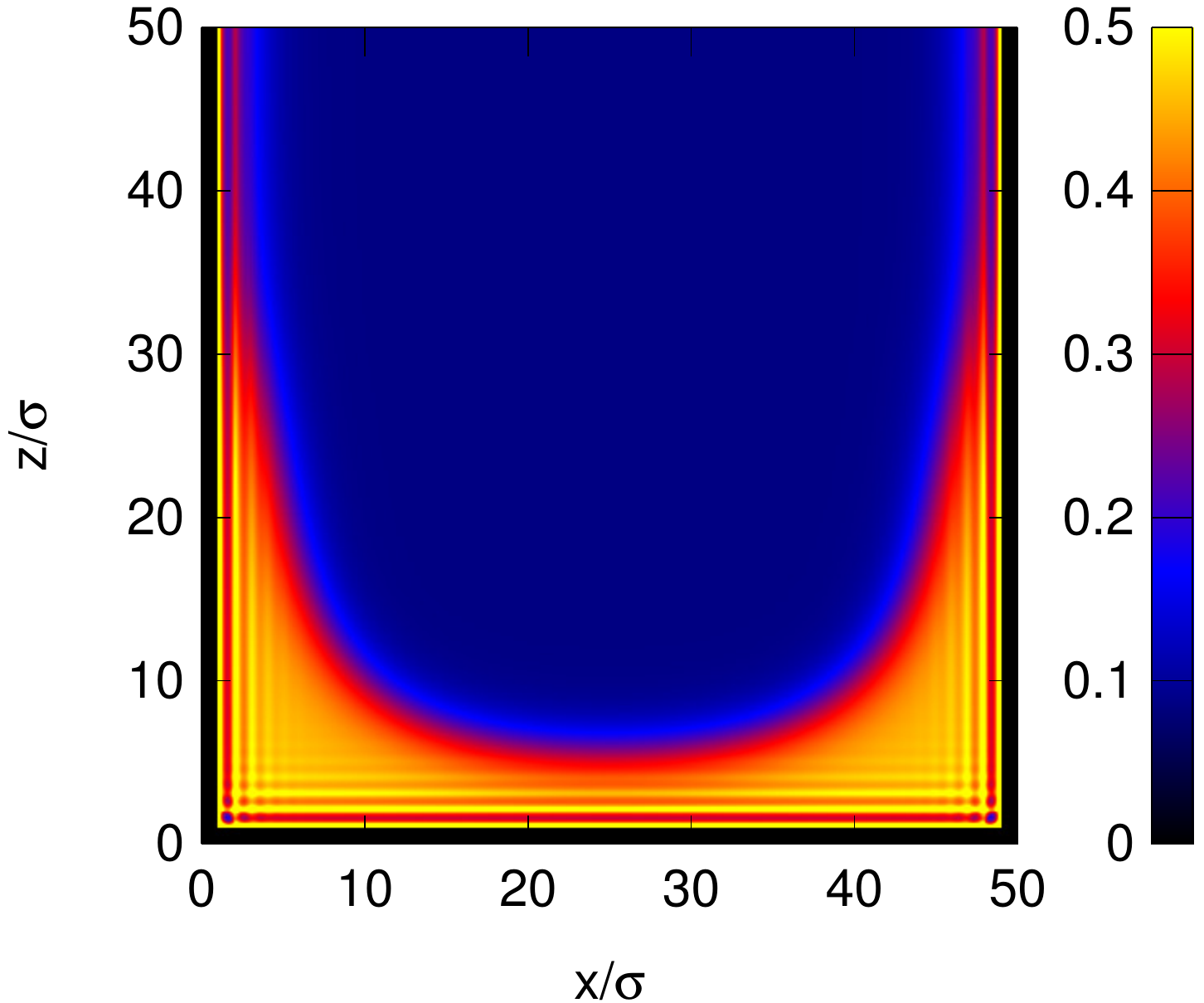} \includegraphics[width=0.22\textwidth]{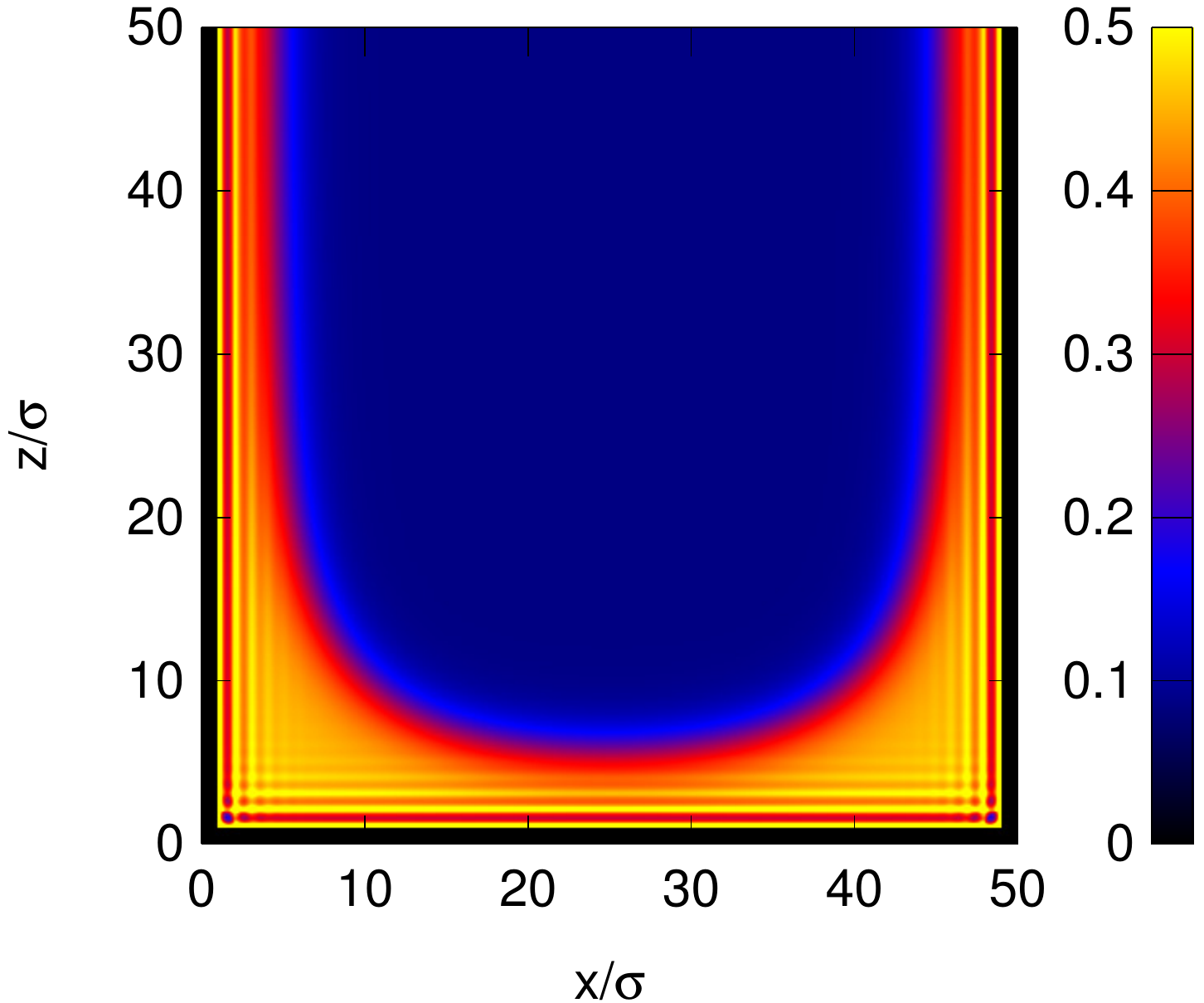}
\caption{Density distributions ($x$-$z$ projection) in a groove with attractive walls at temperature $T=0.91\,T_c$. $L_x=L_z=50\sigma$. Top: coexisting states at a density $\rho_b/\rho_g=0.93695$. Bottom: coexisting states at
a density $\rho_b/\rho_g=0.94813$. }\label{fig5}
\end{figure}

In summary, we have employed a mean-field FMT-DFT to describe adsorption in a single nanogroove with attractive walls that exhibit a first-order wetting
transition. At high temperatures, the results support the picture of the onset of a liquid phase due to a heterogeneous nucleation on
a bottom wall, followed by a continuous rise of the meniscus separating vapour-like and liquid-like phases, and no hysteresis in adsorption appears in this case
(see Fig 1). However, as we have demonstrated, such a scenario is valid only for temperatures above the wetting temperature of a corresponding single wall. Below
$T_w$, the adsorption exhibits a first-order transition (accompanied by hysteresis) as in an open slit.
Thus, $T_w$, which represents a crossover between partial and complete wetting for a single wall, represents a boundary separating first-order and continuous transition regimes for a nanogroove.
Some care is needed, though. For large widths, a wedge filling may take place at the corners of the groove at temperatures above a filling temperature, which may cause a shift of this boundary below $T_w$; for the model
considered here the filling temperature (of a rectangular wedge) is $T_f\approx0.76\,T_c$. We haven't observed such a phenomenon even for a groove width as large as $L_x=50\,\sigma$, but one cannot rule out this possibility.
Finally, it has been shown that in the temperature interval $(T_w, T_{\rm sc})$, the groove-adsorption isotherm may exhibit two loops,
reminiscent of prewetting. The first (if the bottom and side walls are adsorbents of the same strength) corresponds to a jump of the meniscus, which is followed by an discontinuous thickening of the liquid films on the side walls.

We believe that the observed phenomena will motivate further experimental measurements and give some contribution into the long-standing issue of a well-known but
still actively debated phenomenon of hysteretic behaviour of fluid adsorption in mesoporous media.


\begin{thebibliography}{}

\bibitem{gelb1}
L. D. Gelb, K. E. Gubbins, R. Radhakrishnan, and M. Sliwinska-Bartowiak, Rep. Prog. Phys. {\bf 62}, 1573 (1999).

\bibitem{dietrich}
S. Dietrich, in {\it Phase Transitions and Critical Phenomena}, edited by C. Domb and J. L. Lebowitz (Academic, New York, 1988), Vol. 12.


\bibitem{derj}
B. V. Derjaguin, Zh. Fiz. Khim. {\bf 137}, 14 (1940).

\bibitem{evmar}
R. Evans and U. Marini Betollo Marconi, Chem. Phys. Lett. {\bf 114}, 415 (1985).

\bibitem{bonn}
D. Bonn, J. Eggers, J. Indekeu, J. Meunier, and E. Rolley, Rev. Mod. Phys. {\bf 81}, 739 (2009).


\bibitem{nature}
C. Rasc\'on and A. O. Parry, Nature {\bf 407}, 986 (2000).

\bibitem{fil}
A. O. Parry, C. Rasc\'on and A. J. Wood, Phys. Rev. Lett. {\bf 85}, 345 (2000).

\bibitem{fil2}
M. Napi\'orkowski, W. Koch, ans S. Dietrich, Phys. Rev. A {\bf 45}, 5760 (1992).


\bibitem{cov}
C. Rasc\'on and A. O. Parry, Phys. Rev. Lett. {\bf 94}, 096103 (2005).
%
\bibitem{tas}
M. Tasinkevych and S. Dietrich, Phys. Rev, Lett. {\bf 97}, 106102 (2006).


\bibitem{marconi}
U. Marinii Betollo Marconi and F. Van Swol, Phys. Rev. A {\bf 39}, 4109 (1989).

\bibitem{darbellay}
G. A. Darbellay and J. M. Yeomans, J. Phys. A {\bf 25}, 4275 (1992).

\bibitem{evans_cc}
C. Rasc\'on, A. O. Parry, N. B. Wilding, and R. Evans, Phys. Rev. Lett. {\bf 98}, 226101 (2007).

\bibitem{mistura}
L. Bruschi and G. Mistura, J. Low Temp. Phys. {\bf 157}, 206 (2009);

\bibitem{hofmann}
T. Hofmann, M. Tasinkevych, A. Checco, E. Dobisz, S. Dietrich, and B. M. Ocko, Phys. Rev, Lett. {\bf 104}, 106102 (2010).

\bibitem{microfluidics}
H. Bruss, {\it Theoretical microfluidics}, Oxford University Press, Oxford (2008).

\bibitem{gelb}
L. D. Gelb, Mol. Phys. {\bf 100}, 2049 (2002).

\bibitem{roth}
R. Roth and A. O. Parry, Mol. Phys. {\bf 109}, 1159 (2011).


\bibitem{evans_dft}
R. Evans, Adv. Phys. {\bf 28}, 143 (1979).

\bibitem{ros}
Y. Rosenfeld, Phys. Rev. Lett. {\bf 63}, 980 (1989).

\bibitem{prew}
R. Evans and U. Marini Betollo Marconi, Phys. Rev. A {\bf 32}, 3817 (1985).


\end{thebibliography}
\end{document}